\documentclass{article}
\usepackage{hiph-art}
\volnumber{19} \issuenumber{1} \edyear{2005}                             
\frompage{000} \topage{000}                                              
\recrevdate{12 September 2005}                                              
\def\rightmark{Rapidity equilibration}
\def\leftmark{G. Wolschin} 
\title{Rapidity equilibration in d + Au and Au + Au systems} 
\authors{ 
{Georg Wolschin  %
\index{Wolschin, G.} 
}\\[2.812mm]
{\normalsize
\hspace*{-4pt} Institut f\"ur Theoretische Physik der Universit\"at, 
D-69120 Heidelberg, Germany\\[0.2ex] 
}} 
\abstract{In a Relativistic Diffusion Model (RDM), the evolution 
of net-proton rapidity 
spectra with $\sqrt{s_{NN}}$ in 
heavy systems is proposed as an indicator 
for local equilibration and longitudinal expansion.
The broad midrapidity valley recently discovered at RHIC 
in central Au + Au collisions at $\sqrt{s_{NN}}$= 200 GeV suggests
rapid local equilibration which is most likely due to deconfinement,
and fast longitudinal expansion.
Rapidity spectra of produced charged hadrons in d + Au and Au + Au
systems at RHIC energies and their centrality dependence are well 
described
in a three-sources RDM. In central collisions, about 19{\%} of the
produced particles are in the equilibrated midrapidity region for 
d + Au.}
\makeindex
\begin{document}
\maketitle
\section{Introduction}\label{intro}
The production and identification of a transient quark-gluon plasma
in thermal equilibrium is of basic importance in relativistic
heavy-ion physics. In this contribution I propose 
nonequilibrium-statistical 
methods to investigate analytically the gradual thermalization occuring 
in the
course of stopping and particle production at the highest available 
energies. The
approach is tailored to identify the fraction of net baryons, as well 
as of 
produced charged hadrons that attain local thermal equilibrium from 
their 
distribution functions in rapidity or pseudorapidity.
This yields indirect evidence for the extent and
system-size dependence of a locally equilibrated parton plasma.

Net-baryon rapidity distributions are 
sensitive indicators for local equilibration and 
deconfinement in relativistic heavy-ion collisions \cite{wol03,wol04}.
At AGS, SPS and RHIC energies these are clearly nonequilibrium 
distributions 
even in central collisions \cite{wol04}, and also the longitudinal 
distributions of produced particles are not fully thermalized in 
heavy symmetric \cite{biy04} as well as in light asymmetric systems 
\cite{wbs05}.  

To account for the nonequilibrium behavior of the system,
the evolution of net-proton rapidity spectra with 
incident energy $\sqrt{s_{NN}}$= 4.9 to 200 GeV is first studied 
analytically in a Relativistic Diffusion Model (RDM, chapter 2), and
compared to AGS \cite{ahl99}, SPS \cite{app99} and RHIC \cite{bea04} 
central collision data, chapter 3. In addition to the 
nonequilibrium-statistical evolution
as described in the RDM, collective longitudinal expansion
is considered through a comparison of the analytical RDM-solutions 
with the data. It turns out that a three-sources RDM with a small, but
sizeable (14\%) midrapidity source is required from the RHIC net-proton data.

The midrapidity source is more important for produced charged 
hadrons. In central d + Au collisions at RHIC energies,
almost 20 \% of the produced particles are in the equilibrium 
source, chapter 4. Here the centrality dependence is also 
investigated in detail. The fraction of produced 
particles in the midrapidity source tends to be larger in 
central Au + Au collsions \cite{biy04}.
The conclusions are drawn in chapter 5. 
\section{Relativistic Diffusion Model}\label{rdm}
The simplified model (\cite{wol04}, and references therein) as discussed 
in this contribution is based on a linear  
Fokker-Planck equation (FPE) 
for the components $k$ of the distribution function R(y,t) in rapidity 
space, 
$y=0.5\cdot ln((E+p)/(E-p))$,
\begin{equation}
\frac{\partial}{\partial t}R_{k}(y,t)=
\frac{1}{\tau_{y}}\frac{\partial}
{\partial y}\Bigl[(y-y_{eq})\cdot R_{k}(y,t)\Bigr]
+\frac{\partial^2}{\partial^{2} y}\Bigl[D_{y}^{k}
\cdot R_{k}(y,t)\Bigr].
\label{fpe}
\end{equation}\\
The diagonal components $D_{y}^{k}$ of the diffusion tensor  
contain the microscopic
physics in the respective projectile-like (k=1), target-like (k=2)
and central (k=3) regions. They 
account for the broadening of the distribution 
functions through interactions and particle creations. 
In the present investigation the off-diagonal terms of the
diffusion tensor are assumed to be zero.
The rapidity relaxation time $\tau_{y}$ determines
the speed of the statistical equilibration in y-space.

As time goes to infinity, the mean values of the
solutions of Eqs. (\ref{fpe}) approach the equilibrium value $y_{eq}$.
This value is zero for symmetric systems, but for asymmetric systems 
like d + Au it deviates from zero.
We determine it 
from energy- and momentum conservation \cite{bha53,nag84}
in the system of target- and projectile-like participants and hence, it 
depends on impact parameter. This dependence is decisive 
for a detailed description of the measured rapidity
distributions in asymmetric systems:

\begin{equation}
y_{eq}(b)=1/2\cdot ln\frac{<m_{1}^{T}(b)>exp(y_{max})+<m_{2}^{T}(b)>
exp(-y_{max})}
{<m_{2}^{T}(b)>exp(y_{max})+<m_{1}^{T}(b)>exp(-y_{max})}
\label{yeq}
\end{equation}\\
with the beam rapidities y$_{b} = \pm y_{max}$, the transverse
masses $<m_{1,2}^{T}(b)>=\\
\sqrt(m_{1,2}^2(b)+<p_{T}>^2)$, and masses
m$_{1,2}(b)$ of the respective participants 
that depend on the impact parameter $b$. The average 
numbers of participants $N_{1,2}(b)$
in the incident nuclei are calculated from the
geometrical overlap, or from Glauber theory.

The RDM describes the drift towards $y_{eq}$ in a statistical sense.
Whether the mean values of the distribution functions $R_{1}$ and 
$R_{2}$ actually attain $y_{eq}$ depends on the interaction time 
$\tau_{int}$ (the time the system interacts strongly, or the integration time
of (\ref{fpe})). It can be determined from dynamical models or
from parametrizations of two-particle correlation measurements. For
central Au + Au at 200 A GeV, this yields about
$\tau_{int}\simeq 10 fm/c$, which is too short for $R_{1}$ and 
$R_{2}$ in order to reach equilibrium. Note, however, that this does
not apply to $R_{3}$ which is born near local equilibrium at short 
times, and then spreads in time through interactions with other
particles at nearly the same rapidity. 

The analytical diffusion model is consistent with, and 
complementary to parton cascade models
where stopping involves large sudden jumps in rapidity from hard
scatterings (eg. \cite{bas03}), because even hard partons can
participate significantly in equilibration processes, as is evidenced 
by the high-$p_{T}$ suppression found in Au + Au at RHIC.

Nonlinear effects are not considered here. Their
possible role in the context of relativistic
heavy-ion collisions has been discussed elsewhere 
(\cite{wol04}, and references therein).
These account to some extent for the collective expansion of the
system in $y-$space, which is not included a priori in a statistical
treatment. In the linear model, the expansion is
treated through effective
diffusion coefficients $D_{y}^{eff}$ that are larger than the
theoretical values calculated from the dissipation-fluctuation theorem that 
normally relates $D_{y}$ and $\tau_{y}$ to each other \cite{wol03}.
This relation has been derived from the requirement 
that the stationary solution of Eq.(\ref{fpe}) is equated
with a Gaussian approximation to the thermal equilibrium
distribution in y-space. At fixed incident energy, this weak-coupling 
rapidity diffusion coefficient turns out to be proportional 
to the equilibrium temperature T as 
in the analysis of Brownian motion (Einstein relation)
\begin{equation}
D_{y}\propto \frac{T}{\tau_{y}}.
\label{ein}
\end{equation}
One then deduces the collective expansion velocities
from a comparison between data (yielding $D_{y}^{eff}$) and theoretical result
($D_{y}$), as described in more detail in \cite{wol05}.  

The FPE can be solved analytically in the linear case  
with constant $D_{y}^{k}$. For net-baryon rapidity distributions,
the initial conditions are $\delta$-functions at the
beam rapidities $y_{b}=\pm y_{max}$. However, it has been shown that in
addition there exists a central (k=3, equilibrium) source at RHIC energies
which accounts for about 14{\%} of the net-proton yield in Au + Au
collisions at 200 AGeV \cite{wol03}, and is most likely related to
deconfinement. For produced particles the central source 
turns out to be even more important. 
\section{Net-proton rapidity distributions}\label{nprot}
Comparing the solutions of Eq.(\ref{fpe}) to Au + Au central collision (5 
per cent of the cross section) data at AGS-energies $\sqrt{s_{NN}}$= 
4.9 GeV \cite{ahl99}, it turns out that due to the relatively long 
interaction time $\tau_{int}$ and hence, the large ratio 
$\tau_{int}/\tau_{{y}}\simeq 1.08$, the system is in 
rapidity space very close to thermal equilibrium (dash-dotted curve),
with longitudinal 
collective expansion at $v_{coll}^{||}$=0.49 (Fig.1, upper frame) 
\cite{wol04}.
The bell-shaped experimental distribution is in good agreement with 
the solution of Eq.(\ref{fpe}). The distributions remain bell-shaped 
also at lower energies \cite{ahl99}.
\begin{figure}[htb]
\vspace*{-.4cm}
                 \insertplot{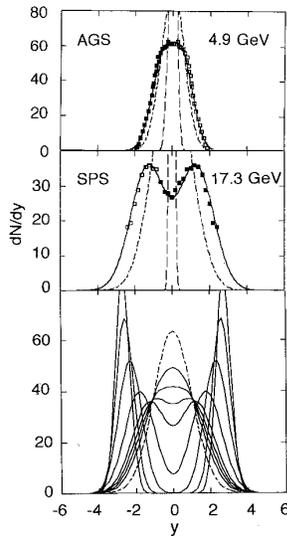}
\vspace*{-1.0cm}
\caption[]{Net-proton rapidity spectra in the two-sources Relativistic Diffusion 
Model (RDM), solid curves, compared to central Au + Au data from AGS, Pb + 
Pb from SPS (middle), and RDM-solutions for various $t/\tau_{{y}}$
(bottom). From \cite{wol05}.}
\label{fig1}
\end{figure}
This situation changes at the higher SPS energy of $\sqrt{s_{NN}}$= 
17.3 GeV (Fig. 1, middle). Here net-proton Pb + Pb rapidity spectra corrected for hyperon 
feeddown \cite{app99} show two pronounced peaks in central collisions, 
which arise from the penetration of the incident baryons through the 
system. The gradual slow-down and broadening is described using Eq.\ref{fpe}
as a hadronic diffusion process with subsequent collective expansion,
$v_{coll}^{||}$=0.75. The associated nonequilibrium 
solutions with expansion for various values of $\tau_{int}/\tau_{{y}}$
are shown in the lower frame of Fig.1. The system clearly does $not$ reach the 
dash-dotted equilibrium solution. Hence, both nonequilibrium 
properties, $and$ collective expansion are required to interpret the
broad rapidity spectra seen at the SPS. 

Within the current framework, no indication for deconfinement of the 
incident baryons or other unusual processes can be deduced from the 
net-proton rapidity data at AGS and SPS energies, because the crucial 
midrapidity region is here too small.

This is, however, different at RHIC energies $\sqrt{s_{NN}}$= 200 
GeV. The RDM nonequilibrium solution exhibits pronounced 
penetration peaks with 
collective longitudinal expansion $v_{coll}^{||}$=0.93, but it fails
to reproduce the BRAHMS net-proton data \cite{bea04} in the broad 
midrapidity valley (solid curve in Fig. 2, bottom): The diffusion of 
the incident baryons due to soft scatterings is not strong enough to
explain the net-baryon density in the central rapidity region.
The individual nonequilibrium solutions $R_{1}$ and $R_{2}$ are Gaussians, and 
if they fit the data points near y=±2, they necessarily grossly
underpredict the midrapidity yield because it is in their tails.
\begin{figure}[htb]
\vspace*{-.4cm}
                 \insertplot{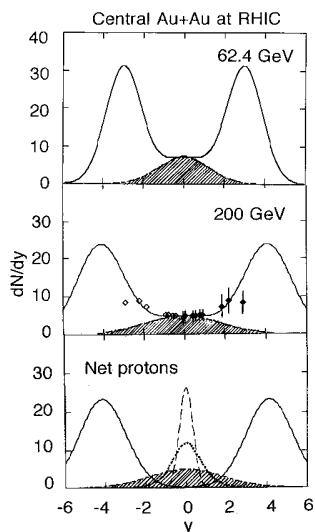}
\vspace*{-1.0cm}
\caption[]{Net-proton rapidity spectra in the three-sources 
Relativistic Diffusion 
Model (RDM), solid curves, compared to central Au + Au data from RHIC.
From \cite{wol05}.}
\label{fig2}
\end{figure}
This central region can only be reached if a fraction of the system undergoes a 
fast transition to local thermal equilibrium, dashed curve in Fig. 2, 
bottom. With collective expansion of this locally equilibrated 
subsystem of 22 net protons ($v_{coll}^{||}$=0.93, dashed areas), the flat 
midrapidity BRAHMS data are well reproduced in an incoherent 
superposition of nonequilibrium and equilibrium solutions of 
(\ref{fpe}) 
\begin{equation}
\frac{dN(y,t=\tau_{int})}{dy}=N_{1}R_{1}(y,\tau_{int})+N_{2}R_{2}(y,\tau_{int})
+N_{eq}R_{eq}^{loc}(y)
\label{normloc}
\end{equation}
with the interaction time $\tau_{int}$ (total integration time of the
differential equation).
The fast transition of a subsystem of $N_{eq}\simeq$55 baryons, or 22 
protons to local thermal equilibrium implies that in this region of
rapidity space, baryon diffusion has been replaced by hard scatterings
of the associated participant partons with large rapidity transfer, 
and subsequent thermal equilibration.

In a schematic calculation for the lower RHIC energy of 
$\sqrt{s_{NN}}$= 62.4 GeV, I have used the rapidity diffusion 
coefficient from 200 GeV to obtain the result in the upper frame of 
Fig. 2.
Two pronounced penetration peaks can be seen, together with a narrow
midrapidity valley. The corresponding data have been taken, and are 
presently being analyzed by the BRAHMS collaboration. 
Once they are available, an adjustment of the effective diffusion 
coefficient and N$_{eq}$ may be required. 

\section{Particle production at RHIC}\label{hauau}
Whereas local equilibration in rapidity space occurs only 
for a small fraction of the
participant baryons in central collisions at RHIC energies, the 
equilibrium fraction tends to be larger for produced hadrons. This can be
inferred from recent applications of the RDM with three sources
(located at the beam rapidities, and at the equilibrium value) and
$\delta$-function initial conditions \cite{biy04,wbs05}. 
After conversion to pseudorapidity space,
the calculations by Biyajima et al. for heavy systems
\cite{biy04} within the
three-sources RDM yield good agreement with Au + Au data at both 130 and 
200 GeV per particle pair. As an example, the result for 130 GeV is 
shown in Fig. 2 in comparison with PHOBOS data; the fit at 200 A 
GeV is of similar quality. 
Recently pseudorapidity
distributions of primary charged particles have become available
\cite{bbb05} as functions of centrality in d + Au collisions at
a nucleon-nucleon center-of-mass energy of 200 GeV. They are also investigated
here within the 3-sources RDM framework.

For produced particles, the initial conditions are not uniquely
defined. Previous experience with the Au + Au system
regarding both net baryons \cite{wol03}, and produced hadrons
\cite{biy04} favors a three-sources 
approach, with $\delta$-function initial conditions at the beam
rapidities, supplemented by a source centered at the equilibrium value
y$_{eq}$. 
\begin{figure}
\vspace*{-.4cm}
        \insertplot{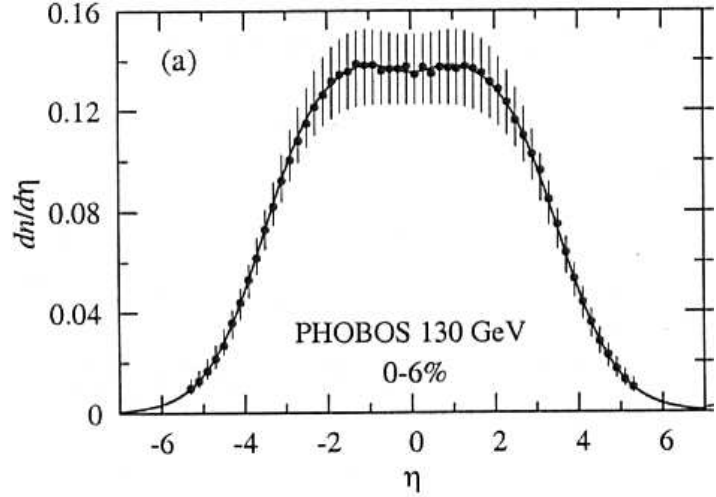}
\vspace*{-1.0cm}
\caption[]{Charged-hadron rapidity spectra in the three-sources 
Relativistic Diffusion 
Model (RDM), solid curves, compared to central Au + Au data from RHIC/PHOBOS.
From Biyajima et al. \cite{biy04}.}
\label{fig3}
\end{figure}
Physically, the particles in this source are expected to be generated
mostly from gluon-gluon collisions since only few valence quarks are
present in the midrapidity region at $\sqrt{s_{NN}}$ = 200 GeV
\cite{wol03}. The final width of this source
corresponds to the local equilibrium temperature of the system which
may approximately be obtained from analyses of particle abundance ratios, plus
the broadening due to the collective expansion of the system.
Formally, the local equilibrium distribution is a solution
of (\ref{fpe}) with diffusion coefficient
$D_{y}^{3}$ = $D_{y}^{eq}$, and $\delta$-function initial condition at the
equilibrium value.
 
The PHOBOS-collaboration has analyzed their minimum-bias data 
successfully using a triple-gaussian fit
\cite{bbb04}. This is consistent with our analytical
three-sources approach, although additional
contributions to particle production have been proposed.
Beyond the precise representation of the data, however,
the Relativistic Diffusion Model offers an analytical description of
the statistical equilibration during the collision and in particular, 
of the extent of the moving midrapidity source which is indicative
of a locally equilibrated parton plasma prior to hadronization.

With $\delta-$function initial conditions for the Au-like source (1),
the d-like source (2) and the equilibrium source (eq), I obtain 
exact analytical diffusion-model solutions as an incoherent
superposition of the distribution functions $R_{k}(y,t)$.
The three individual distributions
are Gaussians with mean values
\begin{equation}
<y_{1,2}(t)>=y_{eq}[1-exp(-t/\tau_{y})] \mp y_{max}\exp{(-t/\tau_{y})}
\label{mean}
\end{equation}
for the sources (1) and (2), and $y_{eq}$ for the
moving equilibrium
source. Hence, all three mean values attain y$_{eq}(b)$ as determined
from (\ref{yeq}) for t$\rightarrow \infty$, whereas for short times
the mean rapidities are smaller than, but close to the Au- and
d-like (absolute) values in the sources 1 and 2. The variances are
\begin{equation}
\sigma_{1,2,eq}^{2}(t)=D_{y}^{1,2,eq}\tau_{y}[1-\exp(-2t/\tau_{y})].
\label{var}
\end{equation}

The charged-particle distribution in rapidity space is then obtained
as incoherent 
superposition of nonequilibrium and local equilibrium solutions of
 (\ref{fpe}) as in (\ref{normloc}), with $N_{k}$ replaced by 
 $N_{ch}^{k}$.
 The average numbers of charged particles in
the Au- and d-like regions $N_{ch}^{1,2}$ are proportional to the respective
numbers of participants $N_{1,2}$,
\begin{equation}
N_{ch}^{1,2}=N_{1,2}\frac{(N_{ch}^{tot}-N_{ch}^{eq})}{(N_{1}+N_{2})}
\label{nch}
\end{equation}
with the constraint $N_{ch}^{tot}$ = $N_{ch}^1$ + $N_{ch}^{2}$ +
$N_{ch}^{eq}$.
Here the total number of charged particles in each centrality bin
$N_{ch}^{tot}$ is determined from the data. The average number
of charged particles in the equilibrium source $N_{ch}^{eq}$ is a
free parameter that is optimized together with the variances
and $\tau_{int}/\tau_{y}$ in a $\chi^{2}$-fit of the data
using the CERN minuit-code.

The results for $N_{1,2}$ in a geometrical overlap calculation
are consistent with the Glauber
calculations reported in \cite{bbb05} which we use in the further
analysis. The corresponding equilibrium values of the rapidity
vary from y$_{eq}=$ - 0.169 for peripheral (80-100$\%$) to 
y$_{eq}=$ - 0.944 for central (0-20$\%$) collisions.
They are negative due to the net longitudinal momentum of the
participants in the laboratory frame, and their absolute
magnitudes decrease with impact parameter since the number of
participants decreases for more peripheral collisions.

The result of the RDM calculation is shown in Fig. 4 (right column, 
middle)
for central collisions (0-20\%) of d + Au. The charged-particle yield is
dominated here by hadrons produced from the Au-like source, but there
is a sizeable equilibrium source (dash-dotted) that is more important
than the d-like contribution. This thermalized source is moving since
y$_{eq}$ has a finite negative value for d + Au, whereas it is at
rest for symmetric systems.

\begin{figure}[htb]
\vspace*{-.4cm}
                 \insertplot{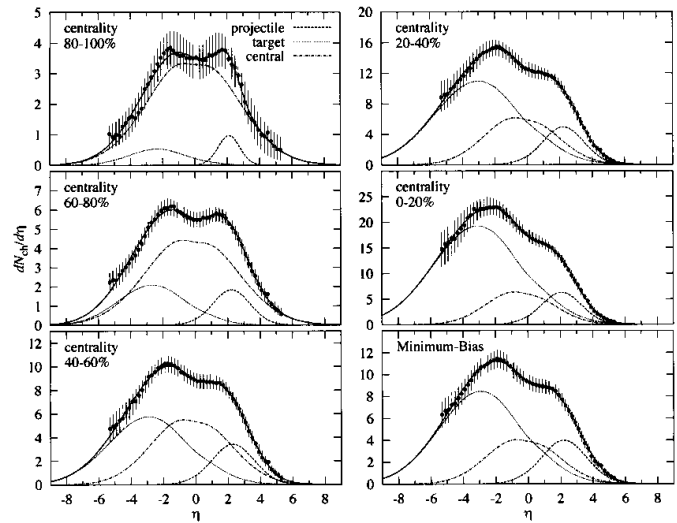}
\vspace*{-1.0cm}
\caption[]{Charged-hadron rapidity spectra in the three-sources 
Relativistic Diffusion 
Model (RDM), solid curves, compared to d + Au data from RHIC/PHOBOS
for various centralities. 
From \cite{wols05}.}
\label{fig4}
\end{figure}

The total yield is
compared to PHOBOS data \cite{bbb05} which refer to the
pseudorapidity $\eta=-ln[tan(\theta / 2)]$ since particle 
identification was not available.
As a consequence, there is a small difference to the model result 
\cite{wbs05}
in $y$-space ($y\approx \eta$) which is most pronounced in the 
midrapidity region. It is removed when
the theoretical result is converted to $\eta$-space  
through the Jacobian
\begin{equation}
J(\eta,\langle m\rangle/\langle p_{T}\rangle) 
 = \cosh({\eta})\cdot [1+(\langle m\rangle/\langle p_{T}\rangle)^{2}
+\sinh^{2}(\eta)]^{-1/2}.
\label{jac}
\end{equation}
Here the average mass $<m>$ of produced charged hadrons in the
central region is approximated by the pion mass $m_{\pi}$, and a
mean transverse momentum $<p_{T}>$ = 0.4 GeV/c is used. In the
Au-like region, the average mass is larger due to the
participant protons, but since their number is small compared to the 
number of produced charged hadrons in the d + Au system, the
increase above the pion mass remains small, and
it turns out to have a negligeable effect on the results.

The magnitude of the equilibrium source for 
particle production is
about 19\% in the light and asymmetric d + Au system, whereas it 
tends to be larger in the Au + Au system \cite{biy04}. 

The model calculations are 
compared with PHOBOS data for 
five centrality cuts \cite{bbb05} and minimum bias \cite{bbb04}
in Fig. 4. 
The observed shift of the distributions towards
the Au-like region in more central collisions, and the steeper slope 
in the deuteron direction as compared to the gold direction
appear in the Relativistic Diffusion Model as a
consequence of the gradual approach to equilibrium: The system is
on its way to statistical equilibrium, but it does not fully reach it.

Given the structure of the differential equation
that is used to model the equilibration,
together with the initial conditions
and the constraints imposed by Eqs. ($\ref{yeq})$ and ($\ref{nch}$),
there is no room for substantial modifications of this result.
In particular, changes in the impact-parameter dependence of the mean 
values in Eq. (\ref{mean}) that are not in accordance with 
Eq. (\ref{yeq}) vitiate the precise agreement with the data.

\section{Conclusion}\label{concl}
To conclude, I have interpreted recent results for
central collisions of heavy systems at AGS, SPS and RHIC energies in
a Relativistic Diffusion Model (RDM) for
multiparticle interactions based on the interplay of
nonequilibrium and local equilibrium ("thermal") solutions.
In the linear version of the model, analytical results for the rapidity
distribution of net protons in central collisions have been
obtained and compared to data. 

The enhancement of the diffusion in rapidity
space as opposed to the expectation from the
weak-coupling dissipation-fluctuation theorem has been interpreted as 
collective expansion, and longitudinal expansion velocities
for net protons have been
determined from a comparison between RDM-results and data based on a
relativistic expression for the collective velocity,
cf. \cite{wol05} for details. A prediction for net-proton
rapidity distributions at $\sqrt{s_{NN}}$= 62.4 GeV yields 
a smaller midrapidity valley than at 200 GeV
which will soon be compared with forthcoming data.
 
Charged-particle production in d + Au collisions at
$\sqrt{s_{NN}}$= 200 GeV as function of centrality
has also been investigated
within the framework of my analytically soluble three-sources 
model, and compared to Au + Au. Excellent agreement with 
recent PHOBOS pseudorapidity
distributions has been obtained. For central collisions, a fraction of only 
19\% of the produced particles arises from the locally equilibrated
midrapidity source, whereas this figure tends to be larger in
Au + Au collisions at the same energy. 

At $\sqrt{s_{NN}}$= 200 GeV, the midrapidity source for particle
production must essentially have partonic origin, because the
net-proton rapidity spectra for Au + Au show that the nucleon
(and hence, valence-quark) content in this region of rapidity
space is small \cite{wol03}. Since the source is in local thermal
equilibrium, it is likely that the generating partons form a
thermalized plasma prior to hadronization.

The d + Au results show clearly that only the midrapidity part of the
distribution function comes very close to thermal equilibrium, whereas the
interaction time is too short for the d- and Au-like parts 
to attain the thermal limit. The same is true for the heavy Au + Au 
system at the same energy, but there the precise fraction of
particles produced in the equilibrium source is more
difficult to determine due to the symmetry of the problem.
\section*{Acknowledgment}
I am grateful to M. Biyajima, T. Mizoguchi and N. Suzuki for 
collaborating on particle production in the RDM. The
hospitality of the Faculty
of Sciences at Shinshu University, and financial support by
the Japan Society for the Promotion of Science (JSPS)
was essential for chapter 4 of this work.

\vfill\eject

\begin{thebibliography}{99}
    
\bibitem{wol03}G. Wolschin, {\it Phys. Lett.} {\bf B 569} (2003) 67.

\bibitem{wol04}G. Wolschin, {\it Eur. Phys. J.} {\bf A5} (1999) 85;
{\it Phys. Rev.} {\bf C69} (2004) 024906.

\bibitem{biy04}M. Biyajima, M. Ide, M. Kaneyama, T. Mizoguchi, and N. Suzuki,\\
{\it Prog. Theor. Phys. Suppl.} {\bf 153} (2004) 344.

\bibitem{wbs05}G. Wolschin, M. Biyajima, T. Mizoguchi, and N. Suzuki,\\
{\it Phys. Lett.} {\bf B 633} (2006) 38.

\bibitem{bha53}H.J. Bhabha, {\it Proc. Roy. Soc. (London) }Ê{\bf A219} (1953) 293.

\bibitem{nag84}S. Nagamiya and M. Gyulassy, {\it Adv. Nucl. Phys.} 
{\bf 13} (1984) 201.

\bibitem{bas03}S.A. Bass, B. M\"uller, and D.K. Srivastava, {\it 
Phys. Rev. Lett.} {\bf 91} (2003) 052302.

\bibitem{ahl99}L. Ahle {\it et al.}, E802 Collaboration,
{\it Phys. Rev.} {\bf C60} (1999) 064901;\\
J. Barrette {\it et al.}, E877 Collaboration,
{\it Phys. Rev.} {\bf C62} (2000) 024901;\\
B. B. Back {\it et al.}, E917 Collaboration,
{\it Phys. Rev. Lett.} {\bf 86} (2001) 1970.

\bibitem{app99}H. Appelsh{\"a}user {\it et al.}, NA49 Collaboration,
{\it Phys. Rev. Lett.} {\bf 82} (1999) 2471.

\bibitem{bea04}I. G. Bearden {\it et al.}, BRAHMS Coll., 
{\it Phys. Rev. Lett.} {\bf 93} (2004) 102301.

\bibitem{wol05}G. Wolschin, 
{\it Europhys. Lett.} {\bf 74} (2006) 29.

\bibitem{bbb05}B.B. Back {\it et al.}, 
{\it Phys. Rev.}, {\bf C72} (2005) 031901.

\bibitem{bbb04}B.B. Back, et al., {\it Phys. Rev. Lett.} {\bf 93} (2004) 082301.

\bibitem{wols05}G. Wolschin, M. Biyajima, T. Mizoguchi, and N. 
Suzuki,\\
{\it Annalen Phys. (Leipzig)} {\bf 15} (2006) 369.\\

\end{thebibliography}
\end{document}